\documentclass[aps,pre,reprint,superscriptaddress,footinbib,twocolumn]{revtex4-2}
\usepackage{blindtext}

\usepackage{color}
\usepackage{gensymb}
\usepackage{amsmath}
\usepackage{amssymb}
\usepackage{dsfont}
\usepackage{xcolor}
\usepackage{graphicx}
\usepackage{url}
\usepackage{breakurl}
\usepackage[breaklinks]{hyperref}
\usepackage{booktabs}
\usepackage{multirow}

\definecolor{airforceblue}{rgb}{0.36, 0.54, 0.66}
\definecolor{amber}{rgb}{1.0, 0.49, 0.0}

\begin{document}
\title{Wrinkling and developable cones in centrally confined sheets}
\author{Lucia Stein-Montalvo}
\altaffiliation{Present address: Department of Civil and Environmental Engineering, Princeton University, Princeton, NJ, 08540}
\email{lsmontal@princeton.edu}
\affiliation{Department of Mechanical Engineering, Boston University, Boston, MA, 02215}
\author{Arman Guerra}
\affiliation{Department of Mechanical Engineering, Boston University, Boston, MA, 02215}
\author{Kanani Almeida}
\affiliation{Department of Mechanical Engineering, Boston University, Boston, MA, 02215}
\author{Ousmane Kodio}
\affiliation{Department of Mathematics, Massachusetts Institute of Technology, Cambridge, MA, 02139}
\author{Douglas P. Holmes}
\email{dpholmes@bu.edu}
\affiliation{Department of Mechanical Engineering, Boston University, Boston, MA, 02215}

\begin{abstract}
Thin sheets respond to confinement by smoothly wrinkling, or by focusing stress into small, sharp regions. From engineering to biology, geology, textiles, and art, thin sheets are packed and confined in a wide variety of ways, and yet fundamental questions remain about how stresses focus and patterns form in these structures. Using experiments and molecular dynamics (MD) simulations, we probe the confinement response of circular sheets, flattened in their central region and quasi-statically drawn through a ring. Wrinkles develop in the outer, free region, then are replaced by a truncated cone, which forms in an abrupt transition to stress focusing. We explore how the force associated with this event, and the number of wrinkles, depend on geometry. Additional cones sequentially pattern the sheet, until axisymmetry is recovered in most geometries. The cone size is sensitive to in-plane geometry. We uncover a coarse-grained description of this geometric dependence, which diverges depending on the proximity to the asymptotic d-cone limit, where the clamp size approaches zero. This work contributes to the characterization of general confinement of thin sheets, while broadening the understanding of the d-cone, a fundamental element of stress focusing, as it appears in realistic settings. 
\end{abstract}

\maketitle
	
\section{Introduction} 
Confinement influences the morphology of countless natural and engineered structures encountered in daily life. It can be a source of frustration or danger, such as in the case of ``railway buckling" when portions of train tracks buckle laterally due to heat. On the other hand, confinement can enable functionality, as in robotic grippers that use a suctioned outer membrane to induce jamming of granular media within~\cite{Brown2010}. It can even enhance aesthetics, as in the wrinkly edges of the Cockscomb flower, which result from differential growth. As these examples suggest, the response to confinement differs depending on the geometry of the confined object. For instance, confinement of ``0D" granular media leads to a transition from gas- or fluid-like to solid-like behavior. This can enable structural emergence~\cite{Guerra2021}, or cause flooding and ecological harm when ice becomes trapped in a river constriction~\cite{Beltaos2008} or fjord~\cite{Burton2018}, altering flow. On the other hand, confinement of 1D structures causes bending or buckling~\cite{Domokos1997}, as demonstrated by the clumping of drying mushroom gills~\cite{Guerra2023} and the buckling of microtubules in vesicles~\cite{Fygenson1997}. In 2D sheets, the response to confinement is more complex. A thin sheet prefers to deform by bending only, as stretching costs much more energy for slender structures. However, unlike a beam, material constraints can conspire with boundary conditions to prevent pure bending. In such cases, the sheet will minimize energy by distributing stress, as in the wavy edges of cooked bacon or a gathered skirt, or focusing it into small, sharp regions--namely, stretching ridges~\cite{Lobkovsky1995} and developable cones or \textit{d-cones}~\cite{Pomeau1995,BenAmar1997}--like those in discarded wrapping paper~\cite{Witten2007}. 

\begin{figure*}[t] 
	\centering
	\includegraphics[width=0.8\linewidth]{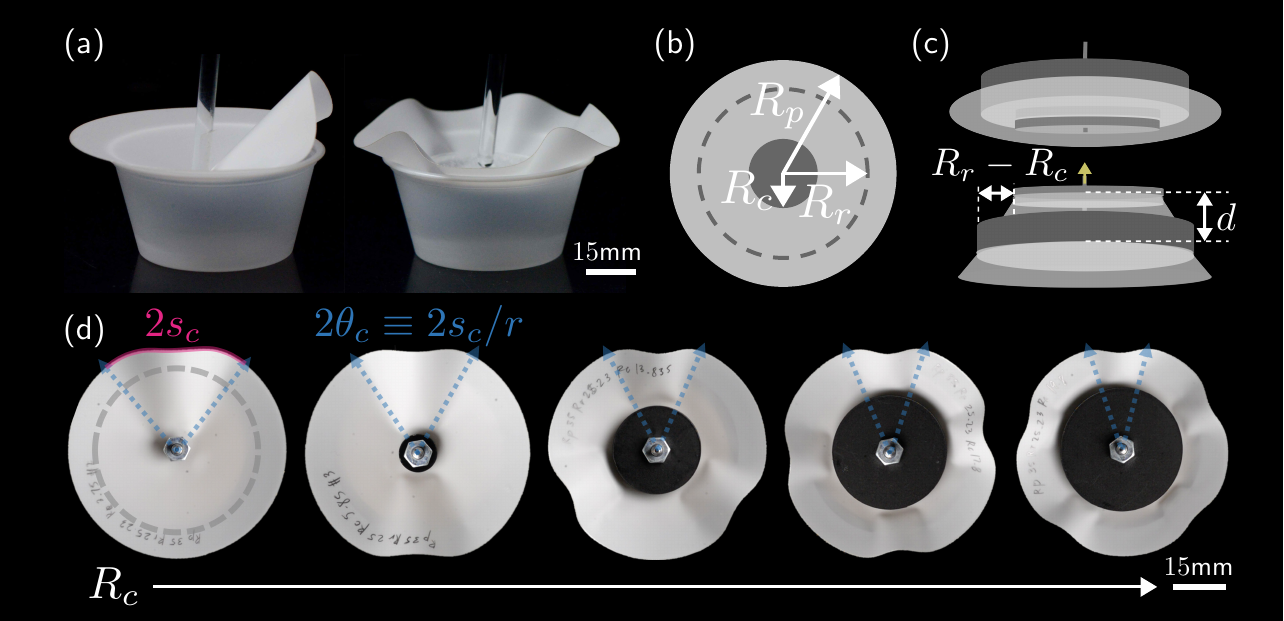}
	\caption{Overview of circumferential buckling in centrally confined, packed sheets. 
		(a) Table-top realizations of the d-cone (left) and our system (right), wherein clamps flatten the central region. 
		(b) Schematic of in-plane parameters: the radii of the clamp $R_c$, ring $R_r$, and plate $R_p$. (Thickness $h$ is not shown.)
		(c) Schematic of experiments. The clamped plate is quasi-statically pulled upward through the ring (green arrow). An upward-facing camera records deformation from below.
		(d) Images from experiments at large $\varepsilon$, when $N=N_\text{max} \in \{1,2,3,4,5\}$. $R_p = 35$ mm, $R_r = 25.2$ mm (dashed gray, first image), and $R_c \in \{3.8,5.9,13.8,17.8,19.7\}$ mm (left to right). The clamp is concealed by the hex nut in the leftmost image. The buckled arc length $2 s_c$ is indicated (at $r=R_p$) in pink (first image). The angular cone size $2 \theta_c(r) = 2 s_c(r)/r$ is marked by the blue arrows, on the flattened sheet. 
	} 
	\label{fig:demo_schematics_Rcsweep}
\end{figure*} 

Developing a theoretical description of how thin sheets deform in response to confinement is challenging, as the Föppl–von Kármán equations that govern the deformations of thin sheets are analytically unsolvable in general. Thus, theoretical progress has been heavily bolstered by experimental and numerical observations, and only certain limits have become well-understood in recent decades. For example, confinement with tensile loads, which can generate compressive stresses due to the Poisson effect, prohibits stress focusing. However, bending comes effectively for free in highly bendable (e.g. ultrathin) sheets~\cite{Davidovitch2011,King2012,Vella2015}, so wrinkles develop instead~\cite{Huang2007,Huang2010,Holmes2010}, their geometry set by compromise between the stretching energy and tensile loads. Wrinkling is also known to emerge when a thin film is compressed while attached to an elastic substrate, e.g. human skin~\cite{Cerda2003,Dillard2018}. The same phenomenon occurs when instead of a physical substrate there is an imaginary one, e.g., fluid weight~\cite{Pocivavsek2008}, curvature~\cite{Paulsen2016,Taffetani2017,Davidovitch2018}, or tension~\cite{Cerda2003,Paulsen2016}, which also provide resistance to large-amplitude bending. 

In the opposing limit, the d-cone, which can be made by pushing a circular sheet through a ring of smaller radius (Fig.~\ref{fig:demo_schematics_Rcsweep}a, left), is stretch-free away from the indenter. Thus, for infinitesimally thin sheets indented with a point-sized indenter, the bulk shape is captured by minimizing the bending contribution to the energy alone, reducing analysis to 1D~\cite{Cerda2005}. This limiting behavior is well-studied from a theoretical perspective. However, unlike many wrinkling scenarios, experimental verification only covers a modest geometric range~\cite{Chaieb1998,Cerda1998,Chaieb1999,Cerda1999}. The transition to stress-focusing, via transient wrinkling, in the d-cone was only very recently explored and rationalized~\cite{Suzanne2022}. Furthermore, debates persist about the size of the stretching core of this fundamental element of thin sheet deformations~\cite{Witten2007,Mowitz2022}. 

Away from these limits, wherein sheets either diffuse or focus stress rather conclusively, intermediate confinement can drive sheets to form coexisting stress-focused and stress-diffuse regions~\cite{Schroll2011,Vandeparre2011}. Furthermore, changing boundary conditions drive spontaneous transitions between these states. These emergent motifs and behaviors are even less understood (linear stability analyses and perturbations to the flat state cannot capture these secondary instabilities) and experiments~\cite{King2012,Timounay2020,Roman2012} and simulations~\cite{Andrejevic2022} have been particularly valuable in a recent surge of effort in this area.



Here, we set out to investigate the intermediate regime of confinement, in which there is a cascading transition of the stress from a diffuse to a focused area. To do so, we examine a system in which we pack a thin, circular sheet into a smaller opening, while forcing a finite-sized region at its center to remain flat (see Fig.~\ref{fig:demo_schematics_Rcsweep}a, right). With a comprehensive experimental study, informed by molecular dynamics (MD) simulations, we address the following question related to thin sheet confinement: How does geometry dictate shape-selection and force response of centrally confined sheets, near and away from the d-cone limit? 

In Sect.~\ref{sect:methods}, we explain our experimental and simulation methods. Then in  Sect.~\ref{sect:wrinklecone}, we describe the rich transitions that occur -- from axisymmetric deformation, to stress-diffuse wrinkling, to stress-focused, sequential formation of truncated cones -- as centrally clamped sheets are pulled through a ring. Next, in Sect.~\ref{sect:fc}, we study the geometric dependence of the critical force at which the transition to stress-focusing occurs, presenting simple rationalizations based on plate buckling, and buckling of a confined ring. In Sect.~\ref{sect:nwthetac}, we present our broad observations of how the cone size, and to a lesser extent, the number of wrinkles, depend on geometry. For cones, we show how this behavior diverges depending on proximity to the asymptotic d-cone limit, i.e. when the clamp size approaches zero, before concluding in Sect.~\ref{sect:discconcl}.



\section{Methods}\label{sect:methods}
In our experiments and simulations, we vary four geometric parameters (see Fig.~\ref{fig:demo_schematics_Rcsweep}c): the plate, or sheet, radius $R_p$, the radius of the ring into which the sheet is packed $R_r$, the radius of the clamp in the center of the sheet, $R_c$, and the sheet thickness $h$. We define a dimensionless packing parameter as $\varepsilon = d/(R_r-R_c)$, where $d$ is the distance the sheet is pulled through the ring. The ranges of parameter combinations are given in Table I. Our experimental and numerical methods are described next.
\begin{table}[h!]
	\label{table}
	\centering
\begin{tabular}{ccc}
	\toprule
	& \multicolumn{1}{c}{Experiments} & \multicolumn{1}{c}{Simulations} \\
	\midrule
	$R_c/R_r$ & $[2.6 \times 10^{-2},9.7 \times 10^{-1}]$ & $(0,5.4 \times 10^{-1}]$ \\
	$R_c/R_p$ & $[1.6 \times 10^{-2}, 8.8\times 10^{-1}]$ & $(0,8.2 \times 10^{-1}]$ \\
	$R_r/R_p$ & $[2.1\times 10^{-1}, 9.5 \times 10^{-1}]$ & $[4.2 \times 10^{-1}, 9.5 \times 10^{-1}]$ \\
	$h/R_p$ & $[1.0 \times 10^{-3}, 2.3\times 10^{-2}]$ & $[2.0 \times 10^{-4}, 9.0 \times 10^{-3}]$ \\
	$h/(R_p-R_c)$ & $[2.0 \times 10^{-3},3.4\times 10^{-2}]$ & $[2.0 \times 10^{-4},1.2 \times 10^{-3}]$ \\
	\bottomrule
\end{tabular}
	\caption{Range of parameter configurations tested in experiments and simulations.}
\end{table}

\subsection{Experiments}
In experiments, plastic sheets of radius $15 \leq R_p \leq 60$ mm
are clamped between pairs of smaller, circular acrylic clamps with radius $1 \leq R_c \leq 36$ mm, and quasi-statically drawn through a ring of radius $12 \leq R_r \leq 52$ mm, where $R_c<R_r<R_p$ (see schematic in Fig.~\ref{fig:demo_schematics_Rcsweep}c.) In the main experiments, we laser-cut (Epilog Laser Helix, 75 W) circular plates of radius $15 \leq R_p \leq 60$ mm from polyethylene terephthalate (PET) sheets (Dupont Teijin Film, McMaster-Carr) with Young's modulus $E = 3.6$ GPa, Poisson's ratio $\nu = 0.38$, and thickness $h = 0.127$ mm. 

\begin{figure}[h!] 
	\centering
	\includegraphics[width=0.7\linewidth]{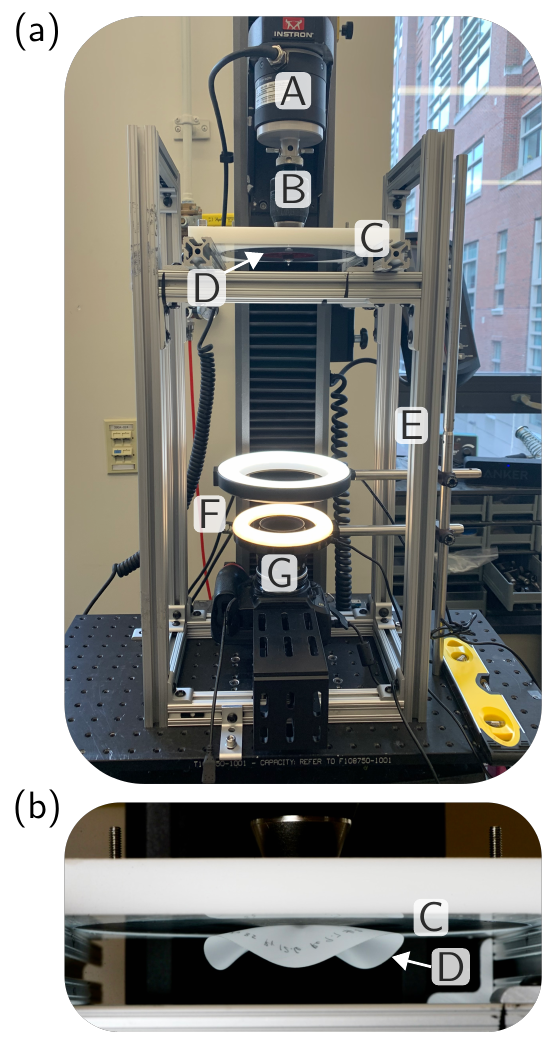}
	\caption[]{Experimental setup. (a) Full setup before an experiment begins, and (b) view from the right as a sheet is pulled through a circular cutout in the stage. Labels correspond to: A: Instron load cell, B: drill-type grip, C: stage with ring cutout, D: clamped sheet, E: aluminum frame, F: light sources, and G: upward-facing camera. 
	} 
	\label{fig:expsetup}
\end{figure}

Prior to cutting, PET sheets were flattened in an oven set to 80$\degree$ C for a minimum of 1 hour, while sandwiched between metal plates. Pairs of circular clamps with radius $1 \leq R_c \leq 36$ mm were cut from acrylic (thickness $6.35$ mm). All plates and clamps were cut with a $3.8$ mm diameter hole at the center so that a $3.6$ mm diameter partially threaded aluminum rod could be fed through the plate, which was sandwiched between two clamps. A hex nut secured the rod-clamp-plate assembly, which was then attached to the $500$ N load cell of the Instron 5943 via a drill-type grip (Instron 0.375 in Keyless Drill-Type Chuck Assembly). A custom-built frame made from T-slotted aluminum rails (McMaster-Carr) was mounted to the base of the Instron (see Fig.~\ref{fig:expsetup}). A CNC-milled polyvinyl chloride (PVC) plate (thickness $9.53$ mm) with a stepped circular cutout (diameter $55$ mm) and three evenly-spaced through-holes was fixed to the top of the frame. Interchangeable PVC rings were screwed into the opening, closing the inner diameter some amount to result in a ring radius $ 12 \leq R_r \leq 52 $ mm (where $R_c < R_r$). For a closer look at small-amplitude wrinkles at low confinement, we performed additional experiments wherein we replaced the opaque PET sheets with reflective colored photo filter film 
($h = 0.075$ mm; Pro Gel, B\&H Photo), and stacked two circular LED ring lights ($4$ inch diameter, B-Qtech, Amazon and $6$ inch diameter, MACTREM, Amazon) concentrically, separated by about 6 inches. Additional experiments to vary the thickness were performed with shim stock ($E = 4.33$ GPa, $\nu = 0.4$,  $0.025 \leq h \leq 0.400 $ mm; Artus). The samples are loaded beyond their elastic limit, so each sheet was used only once.

The plates are very sensitive to initial conditions, so care was taken before each experiment to ensure that the ring and clamped plate were level and centered with respect to one another, and that the clamped plate was just in contact with the underside of the ring. Then, quasi-static displacement-controlled tensile tests were administered using the software Bluehill 3. The clamped plate was pulled upward through the ring at a rate of $0.05-1$ mm/min. Global characteristics like the number of wrinkles and cones, which can be readily observed in general, were recorded during experiments. An upward-facing camera (Nikon D610 DSLR) was also mounted parallel to the clamped sheet and the ring, recording deformation as the imposed upward displacement, $d$, increased. Videos were recorded and used for post-processing alongside Instron force-displacement data. 

The angular cone size $\theta_c$ was measured in complementary experiments. The Instron was stopped and points where the sheet contacts the ring were marked manually, then $\theta_c$ was measured once the sheet was released from the Instron. Force-displacement data was not used for these samples. For several other samples, we took 3D scans (Einscan Pro) of the sheet while the Instron was paused at regular intervals of $d$. In general, we did not observe significant change in $\theta_c$ with $\varepsilon$, as excess length that emerges as packing increases can be accepted by the formation of new cones, as described in Sects.~\ref{sect:wrinklecone} \& ~\ref{sect:thetac}. 

\begin{figure*}[t]
	\centering
	\includegraphics[width=0.9\linewidth]{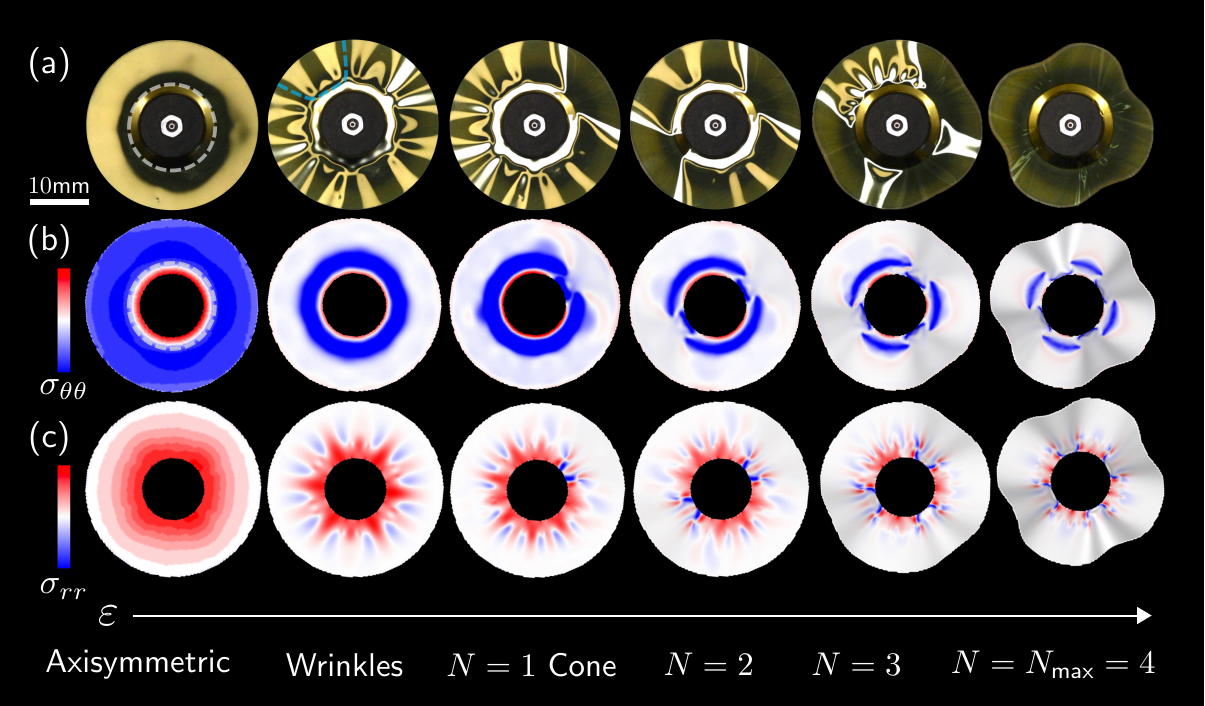}
	\caption{Stages of deformation: axisymmetric, wrinkling, and sequential cone formation. Images from a representative (a) experiment and (b) \& (c) simulation, where $R_p = 30$ mm, $R_r = 15$ mm (dashed gray, first image of (a) \& (b)), $R_c = 11$ mm, and $h=0.075$ mm. As $\varepsilon$ increases, the sheet exhibits a wrinkle-to-cone transition ($n_w=6$; dashed blue curve highlights one wrinkle in (a), second image), and sequential cone formation (up to $N_{\text{max}} = 4$). Color depicts circumferential stress in (b) and radial stress in (c) (Second Piola-Kirkhoff). (Blue: compression, red: tension; Color scale varies between images.) 
	}
	\label{fig:wrinklecone}
\end{figure*} 

To investigate how our clamped boundary conditions compare with indentation, which is the typical boundary condition in d-cone studies, we performed a limited number of additional experiments. For these tests, sheets initially rested above the ring, and were indented at the center with the point of a pencil with a radius of approximately $0.35$ mm, attached to the Instron via the drill-type chuck attachment.  
\begin{figure*}[t] 
	\centering
	\includegraphics[width=0.7\linewidth]{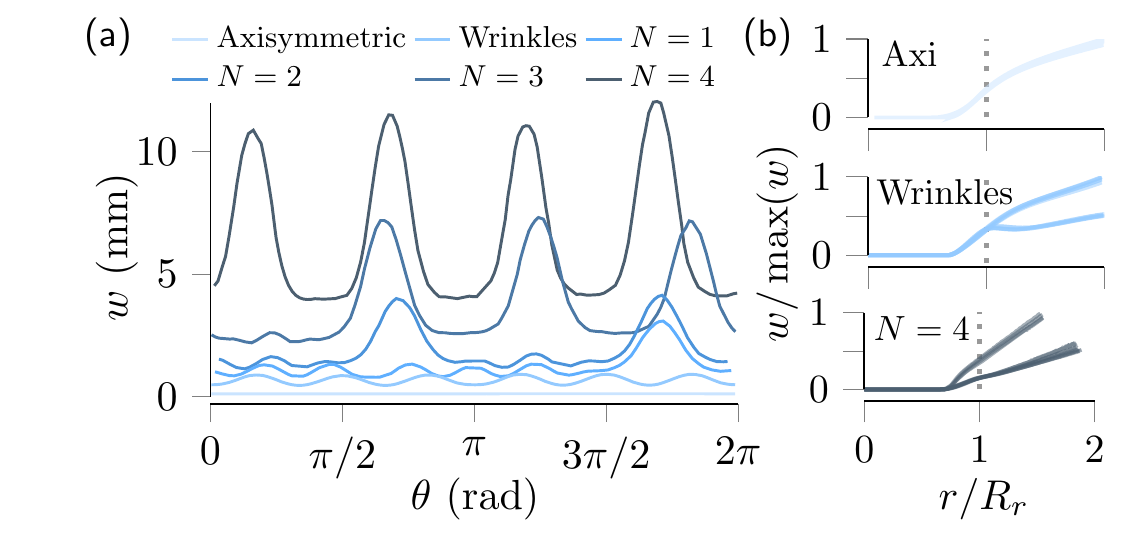}
	\caption{Deformation profiles at different $\varepsilon$.
		(a) Circumferential profiles at $r=2 R_p/3=20$ mm and (b) radial profiles at increments of $\pi/6$,
		corresponding to the simulation in Fig.~\ref{fig:wrinklecone}b \& c ($R_p = 30$ mm, $R_r = 15$ mm, $R_c = 11$ mm, $h=0.075$ mm).
	} 
	\label{fig:profiles}
\end{figure*}
The features we discuss throughout the text, i.e. early behavior, the critical force, and the size of cones, likely do not depend strongly on friction (though sheet self-contact and perhaps when additional cones form may be affected)~\cite{Mellado2011}. Indeed we observe that for fixed geometry, $\theta_c$ is unchanged for a just-cured polyvinylsiloxane (PVS) elastomer sheet (Zhermack Elite Double 32, E = 0.96 MPa), which is much more adhesive than PET. Still, talcum powder coatings and an Anti-Static Gun (Milty Zerostat 3) were used throughout to reduce friction and static charge between the sheet and the ring. 


\subsection{Molecular Dynamics (MD) simulations}
To corroborate our experimental observations and gain insight into features such as the stress distribution (Fig.~\ref{fig:wrinklecone}b\&c), we also performed molecular dynamics (MD) simulations using the large-scale atomic/molecular massively parallel simulator (LAMMPS). Compared to e.g. finite element simulations, MD handles contact well, which is essential for our system. Friction and gravity are absent in simulations. To simulate a plate, we use a triangular lattice of particles, with the potential
\begin{equation}\label{potential}
U_{2d} = \frac{\sqrt{3}}{4} Eh \sum_{ij} (q_{ij}-q_{0})^2 + \frac{Eh^3}{8\sqrt{3}} \sum_{ijk}(1+\cos\theta_{ijk}).
\end{equation}
Here, $q_0$ is the lattice spacing, which is ten times the thickness. The first term adds a harmonic stretching potential between nearest-neighbor particles, and the second term adds a bending potential between all sets of three adjacent, collinear particles. In the limit of small strains compared to unity, and large radii of curvature compared to the lattice spacing, this model is equivalent to an elastic sheet of thickness $h$, Young's modulus $E$, bending rigidity $B = Eh^3/[12(1-\nu^2)]$, and Poisson's ratio $\nu = 1/3$~\cite{Seung1988}. The ring is simulated using a granular pair potential. The plate is offset from the center by a small amount (approximately $0.1$mm, or $0.8h$). Without imposing this asymmetry, the d-cone limit ($R_c \to 0$) results in two cones, which is known to be of similar, but slightly higher, energy than a single cone~\cite{Cerda2005}. We clamp particles by manually enforcing the displacements of all particles within $0\leq r \leq R_c$ to be zero (and to move rigidly in the vertical direction during packing). To achieve the minimum $R_c$ in the clamped d-cone limit, displacement is imposed on the center of a single central grain of radius $0.635$ mm, but the bond can bend through the center of the particle, so $R_c$ approaches $0$. Source code for simulations is provided at~\cite{Guerra2022code}. 

The open source visualization tool OVITO was used alongside custom Matlab and Python scripts for postprocessing, with the force calculated as the derivative of the stretching energy with respect to the bond length between points. The angular size of cones, $\theta_c$, was measured by calculating the angle between particles that contact the ring, and the number of wrinkles were counted by observing the vertical displacements. 

\section{Deformation regimes}\label{sect:wrinklecone}
When the centrally clamped sheet is packed into the boundary set by the smaller ring, the free annulus buckles circumferentially into truncated cones, as shown in Fig.~\ref{fig:demo_schematics_Rcsweep}d. The maximum number of cones, $N_{\text{max}}$, and their characteristic angular size, $\theta_c$, are sensitive to confinement geometry. As the free length $R_p-R_c$ decreases, we see an increase in $N_{\text{max}}$ while $\theta_c$ decreases. The angular cone size $\theta_c$ appears to be much less dependent on the sheet thickness $h$ than the in-plane parameters (see Fig.~\ref{fig:thetac}c). However, this saturated state emerges through a surprisingly rich series of deformation events (see SI movies), which we detail in what follows.

At very small $\varepsilon$, deformation is axisymmetric. Compressive azimuthal stress (i.e. $\sigma_{\theta \theta} < 0$) is felt everywhere outside of a region surrounding the clamp where radial and azimuthal stresses are tensile. However, wrinkles soon relieve compressive azimuthal stresses except in a region just outside of the clamp, which grows with increasing $\varepsilon$, and the tensile core near the clamp remains. Wrinkles are visible in the outer region, i.e. roughly between $R_r$ and $R_p$ (Fig.~\ref{fig:wrinklecone}a\&b). Wrinkling in indented sheets in the d-cone limit ($R_c \to 0$) was first reported from experiments, and rationalized, very recently~\cite{Suzanne2022}, and we confirm this finding in both our experiments and simulations. We observe that wrinkles are evenly distributed about the circumference, and their number, $n_w$, depends on geometry in a similar way to cones, but their wavelength depend less strongly on $R_c$ (see. Fig~\ref{fig:thetac}a.) 

\begin{figure}[h!]
	\centering
	\includegraphics[width=1\linewidth]{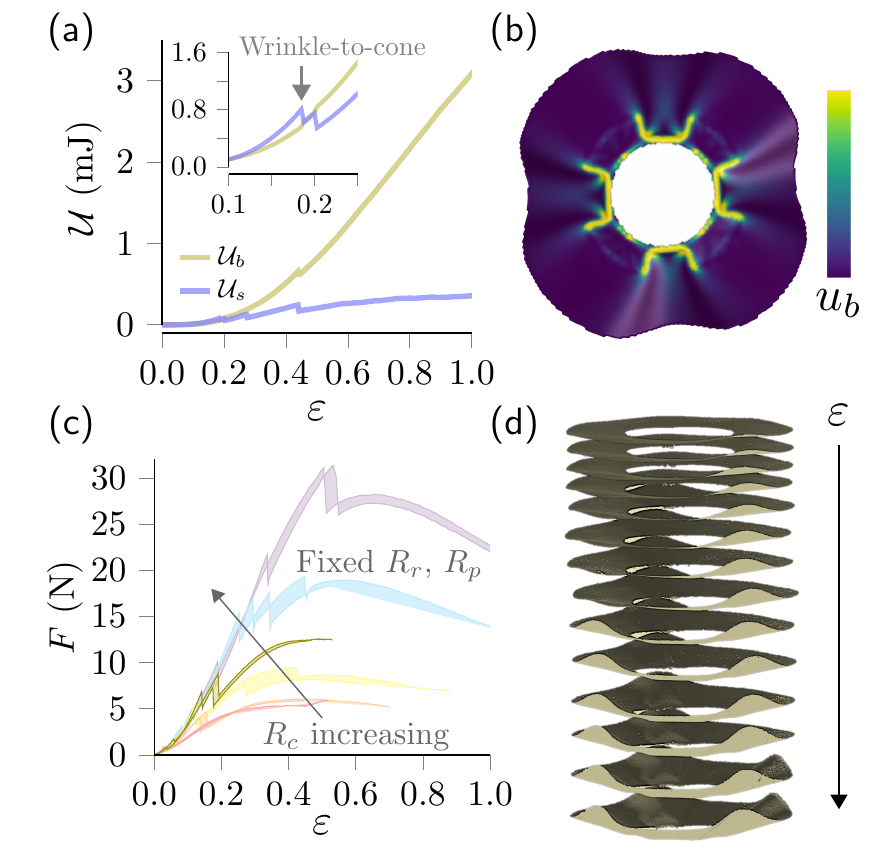}
	\caption{Evolution of energy and force as cones form sequentially.
		(a) Plot of bending and stretching energies for the same simulation as Fig.~\ref{fig:wrinklecone}b \& c ($R_p = 30$ mm, $R_r = 15$ mm, $R_c = 11$ mm, $h=0.075$ mm). Inset plot: Zoomed-in view. 
		(b) Bending energy density at $N_{\text{max}}=4$. 
		(c) Force vs. $\varepsilon$ curves from experiments with $R_p = 35$ mm, $R_r = 25$ mm, $R_c \in \{3, 6, 10, 14, 18, 20\}$ mm, and $h = 0.127$ mm). Two curves are shown for each set of parameters. Drops correspond to cone formation. 
		(d) 3D scans from an experiment with the same geometry as the green curves in (d) ($R_c = 14$ mm).
	}
	\label{fig:en_force}
\end{figure} 

As $\varepsilon$ continues to increase, a truncated cone forms through a sudden buckling event, which is often audible in experiments. The cone accepts enough excess length to replace multiple wrinkles that were in its vicinity, and to reduce the amplitude of, or collapse entirely, any remaining wrinkles. In most cases, the first cone breaks axisymmetry in the stress and deformation fields, however we occasionally observe emergence of multiple cones at indistinguishable $\varepsilon$. Additional cones emerge sequentially and abruptly, and cone formation events are accompanied by sharp drops in the force-displacement curve (Fig.~\ref{fig:en_force}d). Cones may re-arrange as others appear (while deformation is still elastic), until they eventually distribute evenly around the circumference in most geometries. Unlike wrinkles, cones are separated by flat contact lines, reminiscent of the transversely confined elastica~\cite{Chai1998}. The typical shapes of axisymmetric deformations, wrinkles, and cones are compared in Fig.~\ref{fig:profiles}. 

The angular size of cones is set once the first cone appears: as confinement increases, the cone amplitude grows but $\theta_c$ is constant, and additional cones adopt the same size (see Figs.~\ref{fig:wrinklecone}, ~\ref{fig:en_force}d, and ~\ref{fig:thetac}d.) (There are exceptions at high $\varepsilon$ for some geometries, where the structure appears frustrated and will form one or more additional, smaller cone(s).) The bending-dominated elastic energy concentrates in a hinge-like region near the clamp (see Fig.~\ref{fig:en_force}a); the wrinkle-to-cone transition is one from relatively stress-diffuse to stress-focused deformation (Fig.~\ref{fig:wrinklecone}b\&c). 

\begin{figure}[h!] 
	\centering
	\includegraphics[width=\linewidth]{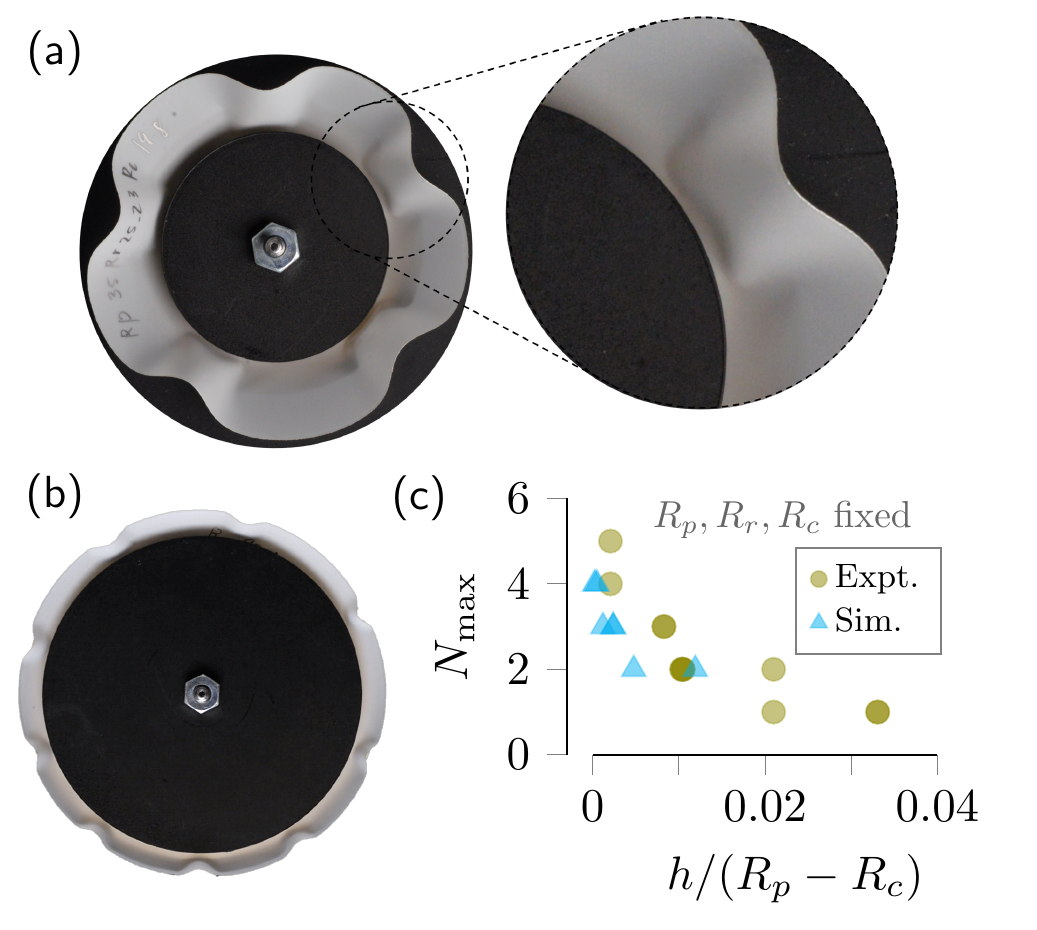}
	\caption{Cone saturation at high $\varepsilon$. (a) Image from experiment showing the development of scars from stress focusing at high $\varepsilon$ in a typical experiment. (b) Image with sparse cones, illustrating how $N_{\text{max}}$ does not necessarily equal $\pi/(2 \theta_c)$. (c) Thickness dependence of $N_{\text{max}}$ for experiments with $R_p = 17.5$ mm, $R_r = 12.6$ mm, $R_c = 5.4$ mm (green dots), and simulations with $R_p = 63.5$ mm, $R_r = 42.3$ mm, $R_c = 10.1$ mm (blue triangles).
	} 
	\label{fig:Nmax}
\end{figure}

When the number of cones saturates at $N=N_{\text{max}}$, increasing $\varepsilon$ causes further focusing. At high values of $\varepsilon$, the endpoints of the concentrated stress region of each conical dislocation near the clamp progressively focus into vertices (see rightmost images in Fig.~\ref{fig:wrinklecone}b\&c); plastic deformation in experiments leads to two sharply curved creases at these endpoints (Fig.~\ref{fig:Nmax}a). This gives the impression that each cone is comprised of two d-cones connected by a straight hinge, which merge in a wrinklon-like manner~\cite{Vandeparre2011} to form a single buckle. We observe that the number of cones is more closely tied to the thickness than is $\theta_c$, suggesting that in-plane stretching matters for setting $N_{\text{max}}$. This thickness dependence, from one set of experiments and one set of simulations with typical geometries, is plotted in Fig.~\ref{fig:Nmax}c. In all stages of deformation, we see from simulations that radial curvature presents in the vicinity of the ring (see Fig.~\ref{fig:profiles}b.), as has been previously reported~\cite{Liang2006,Wang2011}, and thus deformation is not developable.
%

Upon examining the limit $R_c \to 0$, i.e. where the clamp is very small, we note that as for larger $R_c$, the d-cone emerges not through gradual growth, but in an abrupt transition, following wrinkling (see ESI movie 7). This sudden cone formation, accompanied by a drop in the force-displacement curve, has not been reported in previous studies of the d-cone~\cite{Chaieb1998,Chaieb1999}, e.g. with faster~\cite{Mellado2011} displacement rates. 

To summarize our observations broadly, we see that preserving the characteristic size of cones, $\theta_c$, prevails over seeking symmetry, and that both the size of buckled features and the wrinkle-to-cone transition point are closely tied to the in-plane geometric parameters of our system. Understanding how geometry drives pattern formation is central to the study of confinement, and furthermore, the geometric sensitivity we observe differs from the d-cone, whose size is understood to be independent of materials and geometry in ideal sheets~\cite{Cerda2005}. Thus, the two questions we address next are: When does the first cone form, and how does the size of buckled features depend on geometry?

\section{Transition force}\label{sect:fc}
\begin{figure*}[t]
	\centering
	\includegraphics[width=0.7\linewidth]{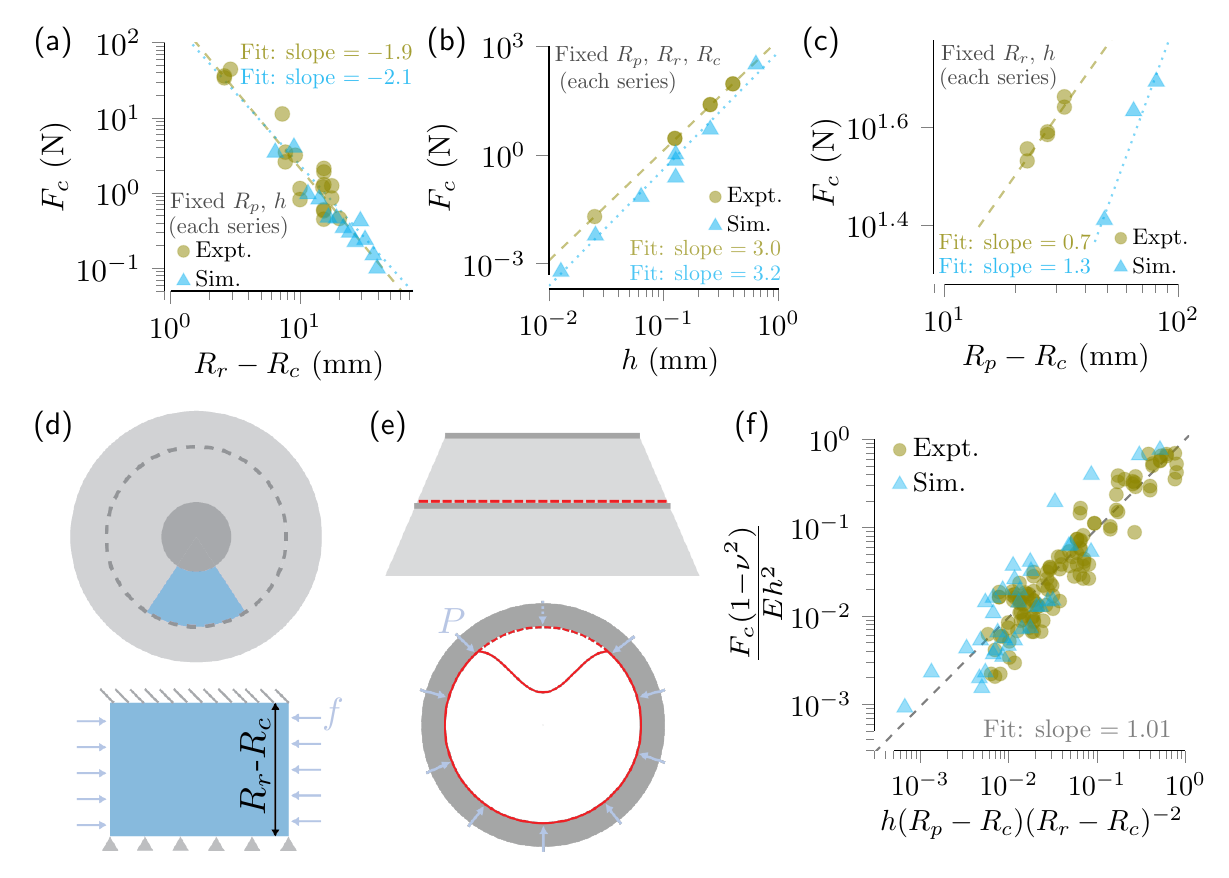}
	\caption{Critical force at which the first cone forms.
		(a) $F_c$ versus $R_r-R_c$ for fixed $R_p$ and $h = 0.127$. Green circles: experiments with $R_p = 35$ mm. Blue triangles: simulations with $R_p = 63.5$ mm. Green, dashed line: log-log fit to experiment series, with a slope of $-1.93$. Blue, dotted line: fit to simulation series, with a slope of $-2.08$. 
		(b) $F_c$ versus $h$ from with fixed $R_p$, $R_r$, and $R_c$. Green circles: experiments with $R_p = 17.5$ mm, $R_r = 12.6$ mm, and $R_c = 5.4$ mm. Blue triangles: simulations with $R_p = 63.5$ mm, $R_r = 42.6$ mm, and $R_c = 10.2$ mm. Green, dashed line: log-log fit to experiment series, with a slope of $3.04$. Blue, dotted line: fit to simulation series, with a slope of $3.23$. 
		(c) $F_c$ versus $R_p-R_c$ for fixed $R_r$ and $h$ (and small $R_r-R_c$). Green circles: experiments with $R_r = 25$ mm ($R_r-R_c = 3.09$). Blue triangles: simulations with $R_r = 45$ mm ($R_r-R_c = 2.6$). Green, dashed line: log-log fit to experiment series, with a slope of $1.30$. Blue, dotted line: fit to simulation series, with a slope of $0.67$. 
		(d) Schematic of the analogy to plate buckling. The section highlighted in blue is compressed, and buckles to form a truncated cone.
		(e) Schematic of the analogy to ring buckling. 	The encased ring (red, radius $R \approx R_r$) is subjected to hydrostatic pressure (blue arrows). Top: side view. Bottom: section view at the ring. Dashed: pre-buckled state.
		(f) A scaling constructed from the empirical findings in a-c and the analogy in d collapses our data, with a fitted slope of $1.008$ (dashed line). 
	}
	\label{fig:force}
\end{figure*}

During wrinkling, compressive hoop stress, which is maximum at $r=R_r$, develops due to excess length. This leads to buckling of a single cone at a critical force $F_c$, which is accompanied by a drop in the force-displacement curve (see Fig.~\ref{fig:en_force}c). Stress focusing -- the energetically preferable deformation mode -- prevails when this critical buckling load is reached. Our experiments and simulations show that $F_c$ scales approximately as $(R_r-R_c)^{-2}$, and as $h^3$, as shown in Fig.~\ref{fig:force} a\&b. These empirical scalings together are likely the dominant length scales for setting the critical force. 

We speculate that the geometric dependence of the critical force could perhaps be rationalized in one of two ways. First, we consider an analogy to plate buckling (see Fig.~\ref{fig:force}d). The critical force per unit length $f_c \sim k B/b^2$, where $B = E h^3/[12(1-\nu^2)]$, $b$ is the width (or radius, for a circular plate) in the direction perpendicular to uniform compressive loading, and $k$ is a dimensionless quantity depending on the plate aspect ratio, boundary conditions, and mode number~\cite{Timoshenko1961}. Setting $b$ to $R_r-R_c$ in our geometry and assuming $k$ is a constant agrees with our empirical scaling, i.e. $F_c (1-\nu^2)/(E h^2) \sim \ell h  (R_r-R_c)^{-2}$. Here, the length scale $\ell$ emerges due to integration of $f$ over the width, i.e. the total force $F_c = f_c \ell$. At small $R_r-R_c$, we infer an approximately linear relationship between $F_c$ and $R_p-R_c$ (Fig.~\ref{fig:force}c). Taking $\ell$ to $R_p-R_c$, we have $F_c (1-\nu^2)/(E h^2) \sim h (R_p-R_c) (R_r-R_c)^{-2}$, which collapses our data as shown in Fig.~\ref{fig:force}f. 

Alternatively, one could imagine that the compressed portion of the annulus which is confined inside the ring is analogous to concentric, rigidly encased, elastic rings subjected to hydrostatic pressure, which causes compressive circumferential stresses to develop (see Fig.~\ref{fig:force}e). (A similar system was also studied theoretically in Ref.~\cite{Cerda2005}, with an emphasis on the analogy to the d-cone, but the onset of buckling is not considered therein.) The critical pressure $P_c$ to buckle a section of a confined ring of radius $R$ is $P_c = E (1-\nu^2)^{-1} (h/2R)^{11/5}$~\cite{Glock1977,Omara1997}. We note that this expression differs from the classical result for the critical pressure of an unconstrained ring, where the critical pressure scales as $[h/(2R)]^3$~\cite{Timoshenko1961}. In the axisymmetric inner region, the material at $r=R_r$ will buckle first as the stress is highest there, so we take $R \to R_r-R_c$. Assuming that the critical force, $F_c$, equals $P_c A$, where $A$ is the cross-sectional area over which the force is distributed, i.e. $A \sim \ell h$ where again, $\ell$ is a length that could reasonably be $R_p-R_c$. This gives, in dimensionless form, $F_c (1-\nu^2)/(E h^2) \sim h^{6/5} (R_p-R_c) (R_r-R_c)^{-11/5}$.

Each of these scalings, which are very similar and in preliminary agreement with our data, relies on the length scale $\ell$, which we have taken to be $R_p-R_c$ according to a limited range of data at small $R_r-R_c$. At larger $R_r-R_c$, however, we observe no clear relationship between $R_p-R_c$ and $F_c$ in experiments (nor for any of the other parameters in our system; see SI for data.) This could signify a sensitivity to imperfections (we note that this limit also corresponds to lower loads, which could amplify this effect due to load cell limitations in experiments), or a change in asymptotic behavior, i.e. the length scale ``felt" by the sheet shifts. As the conical singularity is likely governed by a local scale on the order of the stretching zone, i.e. $R_*$, we speculate that this could be the relevant length when $R_r-R_c$ is large. While these simple analogies may provide a route to a reasonable rationalization for the dependence of the critical force on geometry, there is room for a more precise theoretical treatment.

\section{Wrinkle and cone geometry}\label{sect:nwthetac}
\subsection{Number of wrinkles}\label{sect:nw}
\begin{figure}[h!]
	\centering
	\includegraphics[width=1\linewidth]{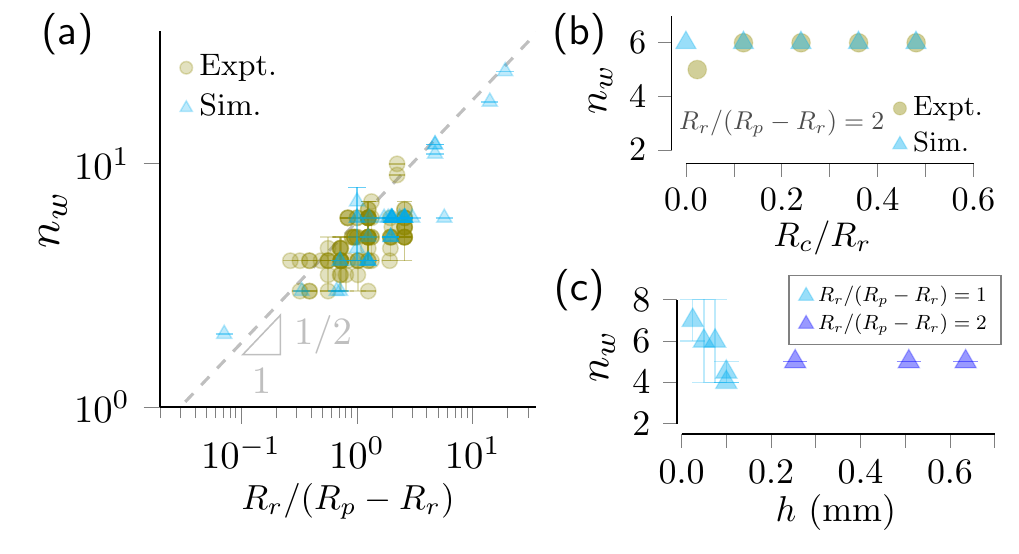}
	\caption{Geometric dependence of the number of wrinkles.
		(a) The dependence of the number of wrinkles $n_w$ on the parameter $R_r/(R_p-R_r)$. The dashed, gray line has slope $1/2$, to guide the eye. Error bars: range of observations as $\varepsilon$ increases. 
		(b) $n_w$ versus $R_c/R_r$ for fixed $R_p$, $R_r$, and $h$ for experiments (green circles) and simulations (blue triangles), showing that $n_w$ is relatively insensitive to $R_c$. 
		(c) $n_w$ versus $h$ for two series of simulations of fixed $R_p$, $R_r$, and $R_c$, with $R_r/(R_p-R_r) = 1$ (light blue triangles), and $R_r/(R_p-R_r) = 2$ (dark blue triangles). The sensitivity of $n_w$ to changing $h$ increases when $h$ is small. 
	}
	\label{fig:nw}
\end{figure} 
Like the critical force, the size and number of cones, and wrinkles -- which precede cones and are evenly distributed about the sheet circumference -- depend on the geometry of the system. Though our focus here is on the size of cones, we first recall that wrinkles emerge in the outer region of the sheet, i.e. between $R_r$ and $R_p$ (Fig.~\ref{fig:profiles}b), and note that the number of wrinkles depends primarily on these radii. The number of wrinkles, $n_w$, versus in-plane geometric parameters is plotted in Fig.~\ref{fig:nw}a. The value of $n_w$ increases with $\varepsilon$ in some cases; observations are averaged in Fig.~\ref{fig:nw}a and error bars represent the range. Though there is a strong quantizing effect of the parameter $n_w$ and clustering around $n_w \leq 6$, or $R_r/(R_p-R_r)\lessapprox 2.5$, we generally observe that the number of wrinkles increases with $R_r/(R_p-R_r)$, i.e. larger wavelengths are preferred when there is more material outside of the ring.
Overall, $n_w \sim [R_r/(R_p-R_r)]^{1/2}$ appears to capture the trend of our data, as shown in Fig.~\ref{fig:nw}a, though this data not cover a wide enough range to be confident in this scaling. We find that the number of wrinkles is relatively insensitive to $R_c$ (see Fig.~\ref{fig:nw}b). When the thickness is small, $n_w$ increases with decreasing $h$, but sensitivity to the thickness decreases for thicker sheets (see Fig.~\ref{fig:nw}c). We find that compared to wrinkles, cones are more sensitive to geometry, and in particular to the size of the clamp, $R_c$, as we examine in detail next.

\subsection{Cone size}\label{sect:thetac}
\begin{figure*}[t!]
	\centering
	\includegraphics[width=0.8\linewidth]{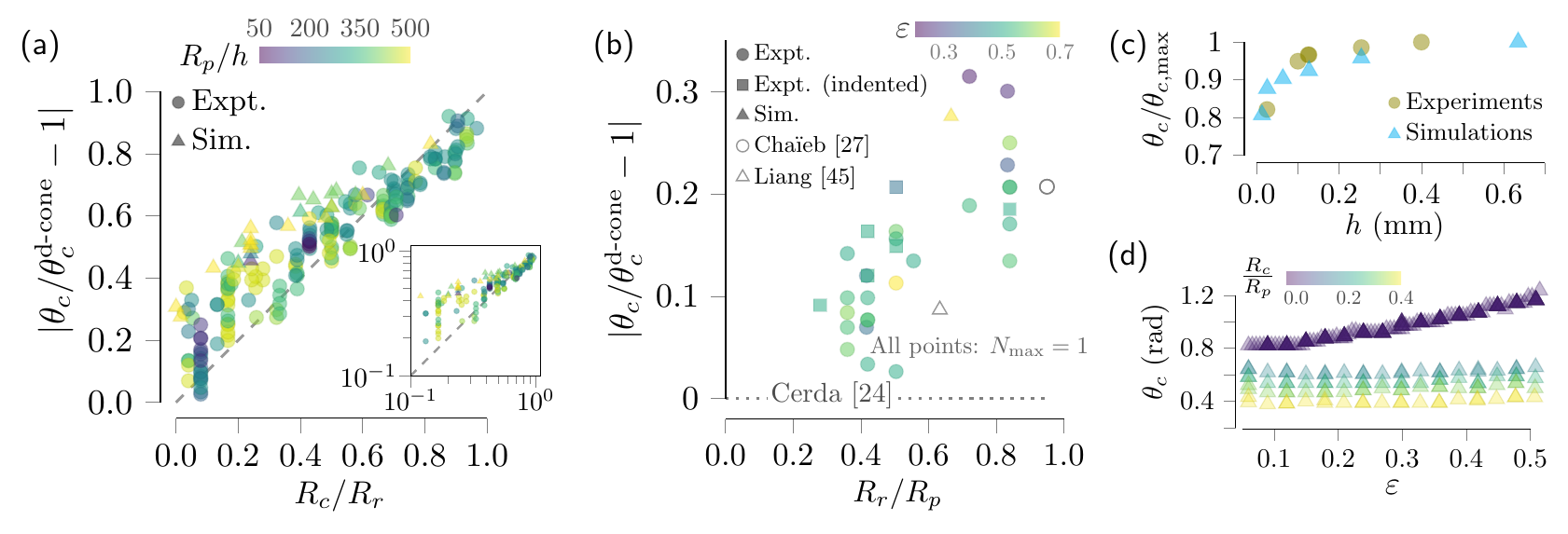}
	\caption{Geometric dependence of the cone size.
		(a) The angular cone size $\theta_c$ normalized by $\theta_c$ for the d-cone ($\theta_c^{\text{d-cone}}=1.21$ rad~\cite{Cerda2005}) decreases with increasing $R_c/R_r$ in general, for relatively large $R_c$. Dotted line: $y=x$. Inset: The same plot in log-log. 
		(b) In the small $R_c$ limit, the cone size is better captured by $R_r/R_p$. Thus, the role of $R_r$ inverts at an intermediate $R_c$. Our data where $N_{\text{max}} = 1$, including experiments where d-cones were produced by indentation instead of clamping, and data from the literature~\cite{Chaieb1999,Liang2005}, are plotted. Data corresponds to $R_c/R_r \lessapprox 0.13$ in (a). Dotted, horizontal line: d-cone solution from~\cite{Cerda2005}.
		(c) $\theta_c$ (normalized by the largest measured $\theta_c$ for each series) versus the sheet thickness. Blue triangles: Simulations with $R_c = 10.16$ mm, $R_r = 42.33$ mm, and $R_p = 63.5$ mm. $\theta_{c,{\text{max}}} = 0.66$ rad. Green circles: Experiments with $R_c = 5.4$ mm, $R_r = 12.6$ mm, and $R_ = 17.5$ mm. $\theta_{c,{\text{max}}} = 0.60$ rad. 
		(d) $\theta_c$ of the first cone that forms versus $\varepsilon$, for $R_p = 50$ mm, $R_r = 33$ mm, and varied $R_c$ in simulations. Away from the small $R_c/R_p$ (single d-cone) limit, $\theta_c$ is unchanged over $\varepsilon$, as additional cones can accept increasing excess length.
	}
	\label{fig:thetac}
\end{figure*} 

The d-cone emerges in the doubly asymptotic limit of $h/R \to 0$ and $R_c/R \to 0$ for characteristic in-plane length scale $R$, where a cone size (at small $\varepsilon$) of $\theta_c^{\text{d-cone}} = 1.21$ rad is predicted to be independent of in-plane geometry~\cite{Cerda2005}. Here, we have examined experiments and simulations that depart significantly from the second asymptotic limit. We find that in general for sheets with finite clamp radius, the angular size of truncated cones is dominated by the relationship between the in-plane parameters $R_c$ and $R_r$. In particular, we observe a general trend of $\theta_c /\theta_c^{\text{d-cone}} \sim 1-R_c/R_r$ over the range of most of our data. This is shown in Fig.~\ref{fig:thetac}a. This trend breaks down in the small $R_c/R$ limit, where we observe increasing scatter in our data. (We also note that $\theta_c$ is systematically slightly lower for simulations than for experiments, which we attribute to the difference in measurement techniques.)

Interestingly, we find that the role of $R_r$ inverts for some intermediate $R_c$. In Fig.~\ref{fig:thetac}b, we plot our data for which $N_{\text{max}} = 1$, which corresponds to $R_c/R_r \leq 0.13$, and we see that the cone size in the d-cone limit follows a general trend of $\theta_c/\theta_c^{\text{d-cone}} \sim 1-R_r/R_p$. We have also included data from experiments with indented d-cones instead of a clamped boundary condition in Fig.~\ref{fig:thetac}b, and available data in the literature~\cite{Chaieb1999,Liang2005}~\footnote{From the simulations of Liang and Witten~\cite{Liang2005}, we take the typical parameters as reported: $R_p = 60$, $R_r = 38$, and $h=0.102$ (lattice spacing of $1$), and $\theta_c \approx 1.105$ rad. In the experiments of Chaïeb et al.~\cite{Chaieb1999}, $h=0.1$ mm, $R_r$ is $5 \%$ smaller than $R_p$ and ranges from $15$ to $90$ mm. To capture this range, we plot these extreme parameter values and their average, with $\theta_c \approx 0.96 \pm 0.04$ rad as reported.} A full understanding of the sensitivity to boundary condition type will require further investigation, but our limited data suggests a relatively weak impact on $\theta_c$. We summarize these diverging, empirical trends in Eq.~\eqref{thetacscalings}, and note that this dependence on in-plane parameters is a significant difference from the theoretical prediction for ideal sheets~\cite{Cerda2005}, likely due to finite thickness in realistic sheets.

\begin{equation} \label{thetacscalings}
\frac{\theta_c}{\theta_c^{\text{d-cone}}} \sim 
    \begin{cases}
 1-\frac{R_r}{R_p} & \text{if } \frac{R_c}{R_r} \to 0 \\
1-\frac{R_c}{R_r}  & \text{if } \frac{R_c}{R_r} \to 1
	\end{cases}
\end{equation} 

We speculate that these empirical trends, which generally capture cone size over our parameter range, emerge due to the following energetic considerations: In the $\frac{R_c}{R_r} \to 0$ limit, the size of the stretching core is believed to scale as $R_* \sim R_r^{2/3} h^{1/3}$~\cite{Witten2007}. The angular size of a d-cone depends on a balance between the elastic energy in the core region and the bending energy in the bulk of the sheet~\cite{Cerda2005}. Thus, we expect that increasing the ring radius $R_r$ raises the core radius, which is perhaps mediated by producing smaller $\theta_c$ so that the core region is confined to a smaller angular extent. 
A large plate radius, $R_p$, offers more area to distribute bending, allowing a larger $\theta_c$. On the other hand, in the relatively large $R_c/R_r$ regime, as the arc length of the concentrated region of high deformation near the clamp scales approximately as $\theta_c R_c$ (see Fig.~\ref{fig:en_force}b.) The sheet seeks to minimize the size of this costly region, so as $R_c$ increases, $\theta_c$ will decrease. Meanwhile, 
since there is also radial bending in regions that contact the ring -- i.e. over a length that scales approximately as $(\pi - \theta_c) R_r$ -- larger cones could reduce the angular extent of this deformation when $R_r$ is large. 
The change in the role of the ring radius in setting $\theta_c$ at intermediate $R_c/R_r$ remains to be explained.

We emphasize that while these low-order trends offer coarse-grained insight, there is much room for refinement. In particular, we do not observe a clear thickness dependence on $\theta_c$ across most of the range of our data, but it likely needs to be accounted for to capture a wider range, as suggested by Fig.~\ref{fig:thetac}c. The in-plane parameters excluded in each limit, i.e. $R_c$ in the small $\frac{R_c}{R_r}$ limit and $R_p$ for large $\frac{R_c}{R_r}$, likely should enter as higher-order terms in a more precise theoretical description. We observe enhanced scatter in the small $\frac{R_c}{R_r} \to 0$ limit, where a single d-cone forms. This scatter may be due to several factors. In this limit, the nature of the core region is likely sensitive to small perturbations to the indenter radius and shape, as well as boundary conditions (clamped versus indented or pinned). This is currently under investigation by the authors~\cite{Stein-Montalvo2023inprep}, but was not studied closely in the present work. Additionally, in contrast to what we observe for most of our data where $N_{\text{max}}>1$, the cone size grows non-negligibly with $\varepsilon$ when only one cone forms, as additional excess length cannot feed other cones (see Fig.~\ref{fig:thetac}c.) It was recently suggested that core size depends on $\varepsilon$ as well~\cite{Suzanne2022}. 

\section{Discussion and conclusions}\label{sect:discconcl}
In summary, we have investigated with experiments and MD simulations the response to confinement of centrally clamped sheets drawn through a ring. We observe a transition from diffuse deformation to stress focusing, in which small-amplitude wrinkles precede the sudden, sequential buckling of truncated cones. We uncover empirical scalings for the force associated with this transition, i.e. $F_c$ scales approximately as $(R_r-R_c)^{-2}$ and the thickness cubed. Two simple models to rationalize the geometric dependence of the force -- based on buckling of a plate and a rigidly encased ring -- produce similar results, and capture our data well despite uncertainty around an ambiguous length scale. In addition to the critical force, confinement geometry also strongly impacts the size of periodic buckled features. Wrinkles, which are visible between the ring and the edge of the plate, depend on the parameter $R_r/(R_p-R_r)$. The angular size of cones at large deformation depends on the geometric parameters of the system in a manner that diverges at intermediate $R_c/R_r$: In the regime where the in-plane geometry approaches that of a d-cone, i.e. relatively small clamp radius, we find empirically that $\theta_c \sim 1-R_r/R_p$. As the clamp size grows, the dependence on the ring radius inverts, and our data is generally captured by the relation $\theta_c \sim 1-R_c/R_r$. 

We are hopeful that these empirical trends could guide future theoretical investigations, which could explain the $R_r$ divergence for $\theta_c$ and offer a more precise description of how the excluded geometric parameters enter at higher order. Another interesting feature of the system is that the number of cones in the saturated state, $N_{\text{max}}$, depends more strongly on the thickness than does $\theta_c$ in our parameter range. This suggests that in-plane stretching contributes to setting $N_{\text{max}}$, but may be negligible in $\theta_c$, and implies a separation of energy scales, which is understood to occur in thin sheets~\cite{Davidovitch2011,Pal2022arxiv}: The dominant energy (bending, in our case) likely determines the macroscale features, whereas subdominant contributions nudge the detailed ones. A complete model to describe the shape-selection in the high $\varepsilon$ regime where cones are present, and establishes greater coherence between the forces and deformations of confined annular sheets, is left to future work. 

The following insights from our findings could have implications for the general confinement and stress focusing in thin sheets in the following ways:
(1) Cones depend strongly $R_c$ but wrinkles do not, highlighting the extreme sensitivity to small geometric changes when stress is focused. 
(2) Relatedly, our data reveal that in realistic sheets with finite length scales, the in-plane geometry dependence of the system departs from the idealized d-cone theory. It remains to determine precisely the role of the type of boundary condition, which was clamped in the main experiments and simulations in this work, as opposed to the indented d-cone~\cite{Stein-Montalvo2023inprep}. However, we expect that this difference would impact shape-selection more when $R_c \approx R_r$, instead of near the small $R_c$ limit.  
(3) Our explanation for when the stress focusing transition occurs may inform other systems that undergo abrupt, secondary instabilities: even when a given deformation mode is energetically preferable, spontaneous transition may be inaccessible until a critical load is reached. Further work would include addressing when subsequent buckling events occur. Our force-displacement data was much less reproducible after $N=1$, suggesting the existence of multiple, energetically comparable configurations.  

Finally, we note that the concentrated region of stress focusing in our model is a perturbed version of the stretching core in the d-cone~\cite{Cerda2005,Liang2005,Witten2007,Mowitz2022}, the size of which remains a conundrum. Though we did not attempt to model this, we hope that our system offers a new window through which to probe this parameter, i.e. by forcing the core region to spread over a larger area. Additionally, our simulations do not include plasticity, which could offer further insight into stress focusing.

\begin{acknowledgments}
LSM, ADG, KA and DPH gratefully acknowledge the financial support from NSF through CMMI-1824882. We also thank Joe Estano in the Engineering Product Innovation Center (EPIC) at Boston University for help with experimental design.
\end{acknowledgments}


%


\end{document}